# Study of Structural Evolution During Controlled Degradation of Ultrathin Polymer Films

**Mojammel H. Mondal and M. Mukherjee**

Surface Physics Division, Saha Institute of Nuclear Physics, 1/AF, Bidhannagar, Kolkata-64, India

ABSTRACT: The structural aspects of polyacrylamide thin films annealed at degradation threshold temperature have been studied as a function of annealing time using in situ X-ray reflectivity technique in vacuum. We observe significant decrease of thickness and increase of density with annealing time for all the films. The dynamical behavior of the changes was modeled in terms of two distinct exponential decay functions, following our earlier observation of two different time scales for the chemical modification pathways, and was found to be in excellent agreement with the data. The diffusion coefficients of the polymer chains corresponding to the two modes are found to be different by an order of magnitude. It was found that the two dynamical modes correspond to the formation of two degradation products at two different rates. The larger time constants for both the modes in case of thickness reduction compare to the chemical changes was explained in terms of inter-chain entanglement and attachment of the polymer with the substrate.

# 1. Introduction

Polymers are widely used in modern industrial societies in a wide range of products such as disposable food service wares, product cases or as structural components in industrially important materials ranging from children's toys to aircraft manufacturing. Environmental factors such as heat, sunlight and chemicals may alter physical and chemical properties of polymer. From both academic and industrial viewpoints it is crucial to understand the mechanism of polymer degradation in terms of their chemical and structural changes. Much research has been devoted to the causes and effects of degradation in polymers[1-13] since the development of the plastic industry. Moreover, a detail understanding of polymers breakdown on heating is important in order to design materials with improved properties and to relate stability and breakdown mechanisms to their chemical structures. In recent years, nanometer scale polymer thin films have drawn tremendous interest due to their technological importance particularly in areas like microelectronics, coatings, biomaterials, membranes, lubricants, adhesives and paints.[14-17] Thin polymer films are interesting because they often exhibit properties that are different from the corresponding bulk polymers therefore the knowledge of the bulk properties is not sufficient in this case. The equilibrium structural and dynamical behaviors of the polymer chains close to the substrate or at interfaces are quite different due to entropic effects and energetic interactions arising at the interfaces. These interfacial effects lead to changes in chain conformation and mobility that affect the properties of the entire polymer film. Response of the long chain polymer molecules at surface and interfaces on annealing at elevated temperature and understanding of related structural changes are therefore of fundamental importance. From a practical point of view it is important to study thermal degradation[18-22] of polymers at the degradation threshold temperature as this temperature determines the upper limit for fabrication when polymers are processed and for their use in applications. It is also important to know the volatile products of degradation in order to guarantee the safety of the workers. Recently we have studied the kinetics of controlled thermal degradation of polyacrylamide ultrathin films by annealing them at the degradation onset temperature (220$^o$C) using X-ray photoelectron spectroscopy (XPS), near-edge X-ray absorption fine structure (NEXAFS) spectroscopy and X-ray reflectivity (XRR) techniques.[1] We observed that the chemical modification of the polymer follows two distinct dynamical modes for the two different final products. It was also observed that despite of chemical modification the interfacial morphology of the film stays almost unmodified indicating that the films retain their polymeric property. However, a study of detail morphological/structural evolution of these ultrathin films during annealing is not available in the literature. This motivates us to perform the present investigation. In the present article we discuss the structural changes of thin polymer films on controlled annealing at degradation onset temperature (220$^o$C). To understand how thermal degradation affects the structural aspects or if there is any influence of the two different degradation modes on the structure modification dynamics of the films during annealing, we have performed X-ray reflectivity measurements with time at this temperature in situ in vacuum. During annealing at 220$^o$C chemical modification of the films occurs and as a result the structures of the films are also changed. It was observed that with annealing time thickness of the films decrease along with increase in their density and reach a saturated value. It was found that thickness reduction with time at 220$^o$C consists of two dynamical modes, much slower compared to their chemical counterpart.

## 2. Experimental

**2.1. Sample Preparation.** In order to understand whether there was any effect of initial polymer concentration or chain morphology, three sets of polymer films were prepared from three different solutions. A powder and an aqueous solution (1%, 10 mg/ml) of high molecular weight polyacrylamide (Supplied by Polysciences, USA) were taken as starting materials for the experiment. Two solutions of concentrations 2 mg/ml and 4 mg/ml were prepared from the powder source (molecular weight 5-6 ×$10^6$). A third solution of concentration 4 mg/ml was prepared from the partially entangled[23] solution source (molecular weight 5 ×$10^6$). The solutions are denoted as solutions A, B and C respectively. Three sets of films from the three solutions were prepared on silicon substrate by spin coating method. The films prepared from the solutions A, B and C were designated as films A, B and C respectively. During the spinning, clean and warm (60°C) air was flown gently over the sol using a homemade arrangement to facilitate faster evaporation of water.[24] Before coating, silicon wafers were cleaned by RCA cleaning method, where the wafers were boiled at 100°C for about 15 minutes in a solution of $H_2O$, $NH_4OH$ and $H_2O_2$ (volume ratio, 2:1:1). The wafers were then rinsed with Millipore water. Apart from cleaning, this treatment enhances the hydrophilicity of the silicon surface by introducing –OH dangling bonds on the surface which helps better attachment of the water soluble polymers. Films of different thicknesses were prepared applying different spinning speeds ranging from 500 to 4000 r.p.m.

**2.2. X-ray Reflectivity.** X-ray reflectivity is one of the best nondestructive methods to measure the thickness, electron density and roughness of ultrathin polymer films. We have used this technique here to study the structural aspects of thermally annealed polyacrylamide films prepared from the three solutions. X-ray reflectivity data were collected in our laboratory setup with CuKα radiation obtained from copper sealed tube anode (2.2 kW, Bruker AXS, D8 Discover). Specular scans with identical incoming and outgoing angles for X-rays were taken as a function of momentum transfer vector $q$ normal to the surface ($q = (4\pi/\lambda) \sin\theta$, with $\theta$ equal to the incident and the reflected angles of the X-ray and $\lambda = 1.54$Å, the wavelength of the radiation). Spin coated polyacrylamide films remain at strained configuration in the as deposited condition. All the films were swelled in saturated water vapor condition in a closed chamber at room temperature for twelve hours to release the strain before they were dried and stored in a desiccator. As the films of water soluble polymer are hygroscopic in nature, it was important to remove absorbed water molecules from the films in order to study their actual structure. At first, data for all the films were collected at room temperature in a chamber continuously evacuated by a rotary pump. The corresponding thicknesses of the films were considered as initial film thickness. The films were subsequently heated at 105°C for 45 min in vacuum and the reflectivity data at this temperature were collected in situ in vacuum. The total time at 105°C including that of data acquisition was about three hours for all the films. From X-ray photoemission spectroscopy studies we found that the films heated above 100°C under vacuum for one hour contains nearly no water and can be considered as practically dry for our study. The structural aspects of the dry films were derived from the X-ray reflectivity data taken at 105°C. Corresponding thickness and density of the films were considered as dry thickness and density of the films.

To study the dynamics of change in film thickness it is important to understand the response of the films as a function of annealing at various temperature and time. In order to investigate the change in film thickness with temperature, we have performed X-ray reflectivity study of a particular film as a function of temperature. The film was dried at 105°C under vacuum for about one hour before the study. The film was then annealed in vacuum at various higher temperatures between 110-230°C. After the sample was stabilized at each temperature for 30 min in situ X-ray reflectivity data were taken in vacuum. Similar data at several intermediate temperatures were also taken during cooling of the sample.

To verify whether the change of thickness with temperature after annealing at 220°C was reversible or not, we have performed another study of a film using X-ray reflectivity technique. The detail process of the experiment is described below. At first the temperature of the film was raised to 220°C in vacuum. Successive in situ reflectivity data at a regular interval of time were taken at this temperature for about 15 hours in vacuum. The temperature of the film was then reduced to 140°C and again increased gradually to 220°C. During this second heating cycle the film was annealed at several intermediate temperatures for about 150 min each and in situ reflectivity data were collected at all the temperatures.

To study the effect of controlled degradation as observed in our earlier XPS and NEXAFS study[1] in terms of structural changes, all the dry films were further annealed at the onset degradation temperature 220°C and in-situ reflectivity data at a regular interval of time were taken at this temperature in vacuum to understand the change of structure of the films with annealing time. The $q$ range for the successive data collection during annealing was carefully optimized to accommodate sufficient number of thickness oscillations as well as to avoid large change during data acquisition. Reasonably good statistics were obtained in 10-15 min, during which data were collected for each thickness. After about 12 hrs of annealing when the change was nearly saturated a data was taken up to high *q* value for obtaining accurate structural information of the films.

The reflectivity data were analyzed using Parrat formalism[25] modified to include interfacial roughness [26] to obtain information about the thickness and electron density of the films. For the analysis of the X-ray reflectivity data, the input electron density profiles were divided into several boxes of thickness equal or more than *$2\pi/q_{max}$* and the interfacial roughness were kept within 2-8 Å. During the analysis, the roughness of the polymer surface, the electron density, the thickness of the films and the roughness of the substrate were used as fitting parameters. A typical reflectivity data with the fitted profiles of a film of initial thickness 338Å treated at different temperatures have been shown in figure 1. The corresponding electron density profiles shown in inset of the figure indicate that thickness of the film reduces and its density increases on annealing.

## 3. Results and Discussion

In figure 2 we have shown the thickness of the film as a function of temperature during the heating and the cooling cycle. It can be seen from the figure that the thickness of the film remains nearly constant up to about 160°C. Beyond this temperature there was a reduction of the film thickness. The reduction rate (with respect to change of temperature) was found to be lower up to about 210°C and increase thereafter. This initial "slow" reduction of thickness may be attributed to the compaction of the films due to larger mobility of the polymer chains above the glass transition temperature (165°C for polyacrylamide) as observed earlier.[23, 27] The increase of rate of thickness reduction above 210°C indicates possibility of additional chemical changes in the film. In the cooling cycle there was a small reduction of the film thickness at the initial stage probably due to the inertial effects of the slow chain dynamics, thereafter the thickness of the film remains constant. This shows that the change of thickness here is permanent in nature and not due to any physical reason such as negative thermal expansion.[28]

In figure 3 we have shown the variation of thickness of the film during the long annealing at 220°C along with those during successive cooling and heating cycles. The inset of the figure shows the reduction of the film thickness as a function of time during the first long duration annealing. The heating cycle data from 140°C to 220°C indicates that there was no change of film thickness once the film was pre-annealed at 220°C for a long time. The absence of thickness change during the second heating cycle indicates that there was no radiation damage and the film attains chemical and structural stability after sufficient annealing at 220°C. The lowering of thickness during the cooling of the film from 220°C to 140°C may be attributed to the inertial effects as also shown in figure 2.

During annealing at onset degradation temperature 220°C there was a substantial loss of mass due to the change of chemical composition of the polymer as shown in figure 4. This leads to formation of voids in the film structure. Due to the thermal motion, the polymer chains occupy the voids and as in-plane motion is restricted for the ultrathin films,[24,28] this would result in a reduction of film thickness.

In figure 5 we have shown the thickness of some of the films as a function of annealing time. It can be clearly seen from the data that the thickness of the films reduces to a saturated value with annealing time. The attempt to fit the reduction of film thickness data with a single exponential function was not successful in general. The dynamical behavior of thickness reduction was modeled in terms of the combination of two single exponential decay functions assuming at least two dynamical modes to be present in the system and was found to be in excellent agreement with the data. The parameters obtained from the fitting are tabulated in table 1 for all the films. When averaged over all samples the time constant for the fast mode was found to be 28 ± 17 min whereas for the slow mode this number was 239 ± 95 min. This indicates that the two modes are widely separated in the time scale. Assuming the relation between the time constant $\tau$ and the diffusion constant $D$ of the polymer chains as $\tau = N/2D$, where $N$ represents the degree of polymerization, one can obtain the diffusion coefficients of the polymer chains in the fast and the slow modes as $2.2 \times 10^{-15}$ and $2.5 \times 10^{-16}$ cm$^2$/sec respectively. It was noted from our XPS and NEXAFS study that onset thermal degradation of polyacrylamide films[1] results in two chemical compounds namely monocyclic and bicyclic polyimides at two different rates. In figure x we have shown the dynamics of formation of the two chemical compounds during the onset thermal degradation taken from our previous study. The data have been fitted with single exponential growth functions to estimate the time scales for the individual reactions. It was observed from the fittings that the time constants corresponding to formation of monocyclic imide was ~15 ± 3 min whereas for bicyclic this value was ~40 ± 7 min. It is likely that the voids created during the changes at two different rates are annealed out also through two distinct dynamical process with two different time constants. It may be noted that the chemical changes are instantaneously reflected in the above experiments. On the other hand, as the polymer chains are attached to the substrate and entangled through inter chain entanglement, the time constants corresponding to the reduction of thickness are likely to be larger compared to their chemical counterparts which may explain the larger values of the two time constants observed for thickness reduction.

Apart from the reduction in thickness the electron density (equivalent to mass density) of the films were found to increase for most of the samples as a result of annealing. The average electron densities of the films were calculated by integrating the electron density profiles obtained from the fitting of the reflectivity data (viz. inset of fig.1). In figure 6 we have plotted the average electron density for all the films before and after annealing as a function of spinning speed of their preparation.[23]

In table 1 percentage increase of density values with respect to dry films are tabulated for all the films. The reduction of film thickness for various films was between 8.4 to 16.7% and the increase of density ranges from 0-12.6% (compared to those of the dry films) without having any systematic dependence between the two. In our earlier observation[1] we estimated that thermally modified polyacrylamide converts to a combination of about 60% monocyclic and 40% bicyclic imide moieties which corresponds to a maximum of about 14% mass loss in the process. The mass density calculated by ChemSketch[29] with chains of 30 monomer units of polyacrylamide, monocyclic and bicyclic polyimides were found to be 1.38, 1.39 and 1.93 gm/cm$^3$ respectively. Therefore if there is a formation of bicyclic imide and full conversion occurs, the density of the final material should be 1.6 gm/cm$^3$, an increase of about 16%. If one considers the reduction of thickness due to mass loss and the increase of density together, the thickness of the films should decrease by up to a maximum of 25%. It is interesting to note from table 1 that the thickness reduction for two thin films (83 and 108Å) were highest (16.7%) with a small or no change of electron density where the data could be fitted only with a single exponential decay corresponding to the fast mode. We believe, for these two films, chemical conversion was limited only to the formation of monocyclic imide. As the density of this material is close to that of polyacrylamide the density of the films remains nearly unchanged. For all other films an increase of density was associated with the reduction of film thickness and both fast and slow modes were required to fit the reduction of thickness data. It may be observed from table 1 that the percentage decrease of thickness and increase of density of the material were less than the calculated numbers in most of the cases. The mass loss during the chemical modification was due to the release of water and ammonia molecules which generates a large amount of free space in terms of free volume or pore in the films. These empty spaces in the film structures prohibits the expected reduction of thickness and increase of densities. Removal of free volume from the film structure strongly depends on the mobility of the chains, therefore may be controlled by the limit set by the attachment of the polymer with the substrate and the inter chain entanglement that prevent the free movement of the long polymer molecules. The requirement of the slow mode in addition to the fast for the films that show increase of density indicates that the slow mode was responsible for increasing the density of the films. As the formation of bicyclic imide is slow (fig.7) and it associates with an increase of density, this is believed to be the cause of slow dynamical mode in the thickness reduction process.

**4. Conclusions**

The structural aspects of thermally modified ultrathin polyacrylamide films have been studied using X-ray reflectivity technique at 220°C. The films were prepared using aqueous polyacrylamide solutions of different concentrations by spin coating method. Successive reflectivity measurements were performed in situ at 220°C with time to study the structural aspects of the ultrathin films during thermal modification. It has been observed that the thickness reduces to a saturated value with annealing time at 220°C for all the films. The dynamical behavior of thickness reduction was modeled in terms of the combination of two exponential decay functions in general and was found to be in excellent agreement with the data. The two modes were believed to correspond to the two different rates of formation of degradation products. The larger time constants for both the modes in case of thickness reduction as compared to those of chemical changes along with the lower reduction of thickness and lower value of final density was explained in terms of substrate attachment and inter chain entanglement of the polymer molecules that hinders the free movement of the polymer chains.

**Acknowledgements**

Authors thankfully acknowledge Prof. S. Hazra for extending the X-ray reflectivity facility for the study.

Table Caption:

**Table 1:** Various parameters obtained from the fitting of thickness reduction data for all the films.

**Table 1**

| Samples prepared from | Spin speed (r.p.m) | Dry thickness (Å) | Thickness reductions (%) | | | Time constants (min) | | Density change (%) |
|---|---|---|---|---|---|---|---|---|
| | | | Total | Fast mode | Slow mode | Fast mode | Slow mode | |
| Solution A | 3050 | 64 | 14.3 | 6.7 ± 0.9 | 7.6 ± 0.8 | 19.6 ± 5.6 | 301.1 ± 48.7 | 12.6 |
| | 1020 | 83 | 16.7 | 16.7 ± 0.8 | -- | 36.5 ± 5.5 | -- | 0 |
| | 1000 | 108 | 16.7 | 16.7 ± 0.1 | -- | 26.7 ± 1.1 | -- | 0.2 |
| | 500 | 128 | 12.0 | 4.5 ± 2.1 | 7.5 ± 2.2 | 44.2 ± 26.2 | 241.9 ± 57.1 | 3.1 |
| Solution B | 1650 | 173 | 15.0 | 4.4 ± 0.6 | 10.6 ± 0.5 | 6.9 ± 3.9 | 142.2 ± 10.3 | 8.9 |
| | 1250 | 223 | 11.6 | 2.5 ± 0.6 | 9.1 ± 0.5 | 6.0 ± 7.3 | 120.5 ± 13.9 | 7.0 |

|  | 1040 | 259 | 13.4 | 6.1 ± 0.6 | 7.3 ± 0.5 | 6.2 ± 2.4 | 111.8 ± 14.7 | 11.6 |
|---|---|---|---|---|---|---|---|---|
| Solution C | 4000 | 182 | 13.0 | 7.1 ± 0.8 | 5.9 ± 0.6 | 17.4 ± 4.4 | 193.4 ± 41.7 | 9.1 |
|  | 2000 | 251 | 14.3 | 10.0 ± 0.6 | 4.3 ± 0.6 | 30.3 ± 3.7 | 370.2 ± 63.2 | 5.8 |
|  | 1200 | 330 | 15.1 | 11.3 ± 0.3 | 3.8 ± 0.003 | 25.2 ± 1.8 | 385.8 ± 38.9 | 10.0 |
|  | 800 | 403 | 8.4 | 4.0 ± 0.8 | 4.4 ± 0.8 | 35.0 ± 16.7 | 280.5 ± 47.0 | 4.5 |
|  | 600 | 451 | 13.4 | 9.3 ± 0.01 | 4.1 ± 0.9 | 46.7 ± 6.1 | 288.8 ± 56.1 | 2.3 |
|  | 500 | 570 | 13.0 | 9.5 ± 0.4 | 3.5 ± 0 | 61.3 ± 6.2 | 190.6 ± 58.5 | 10.0 |

Figure captions:

**Figure 1.** X-ray reflectivity data (symbols) with fitted profiles (lines) of as grown film (a), annealed at 105°C (b) and annealed at 220°C (c) of a particular film of initial thickness 338Å. Inset shows corresponding electron density profiles. The values 0.0 and 0.7 Å$^{-3}$ in electron density corresponds to air and silicon substrate respectively.

**Figure 2.** Change of film thickness with temperature. Heating and cooling cycles are shown against the corresponding symbols. The dashed lines are guide to the eyes to show the thickness change behavior at different ranges of temperature.

**Figure 3.** Change of film thickness with temperature. The arrows show the direction in time. Inset shows the change of thickness with annealing time at 220°C.

**Figure 4.** Formation of monocyclic and bicyclic imide during annealing of the polyacrylamide films at 220°C.

**Figure 5.** The thickness of the films as a function of annealing time at 220°C. Initial thicknesses of the films are shown against the symbols. The lines are obtained by fitting the data using a double exponential function. The inset of the figure shows a single exponential fit to the data of 451Å film.

**Figure 6.** Average electron density of the films as a function of spinning speed. The open and solid symbols represent the average electron densities of the films at 105°C and 220°C respectively. The downward triangles, upward triangles and circles represent the average electron densities of the films A, B and C respectively. The error bars in the figure shows the variation of density when the least square fit value in the analysis of the reflectivity data is sacrificed by ± 1%. The two dashed lines in the figure are the exponential fit to the data.

**Figure 7.** The variations of concentrations of monocyclic (blue dots) and bicyclic imide (red squares) compounds with annealing time. The lines are obtained by fitting the data with an exponential growth function, $y = A(1 - e^{-t/\tau})$. The time constants $\tau$ for monocyclic and bicyclic compounds are 15 ± 3 min and 40 ± 7 min, respectively.

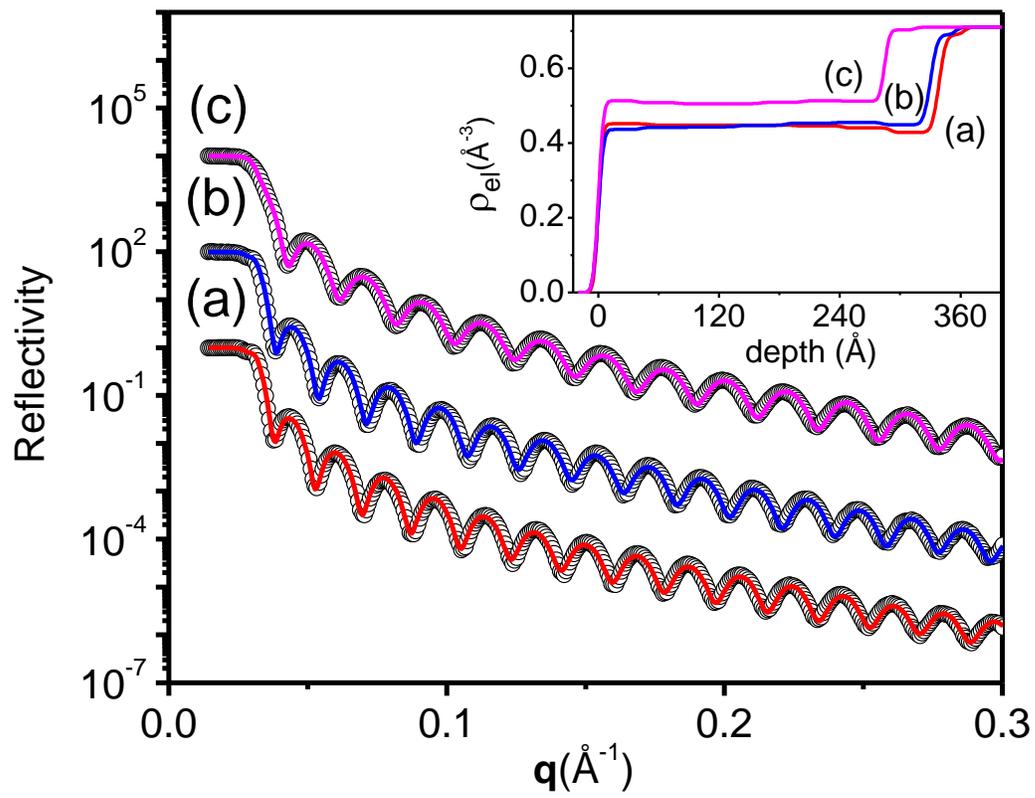

**Figure 1.**

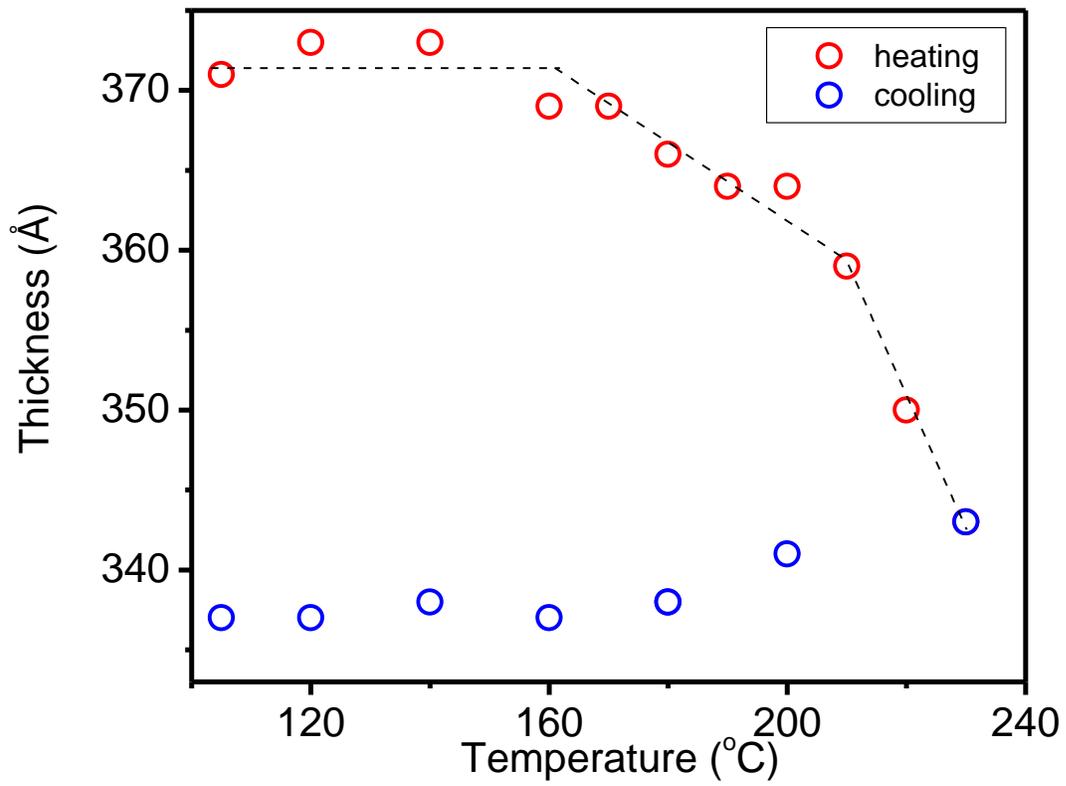

**Figure 2.**

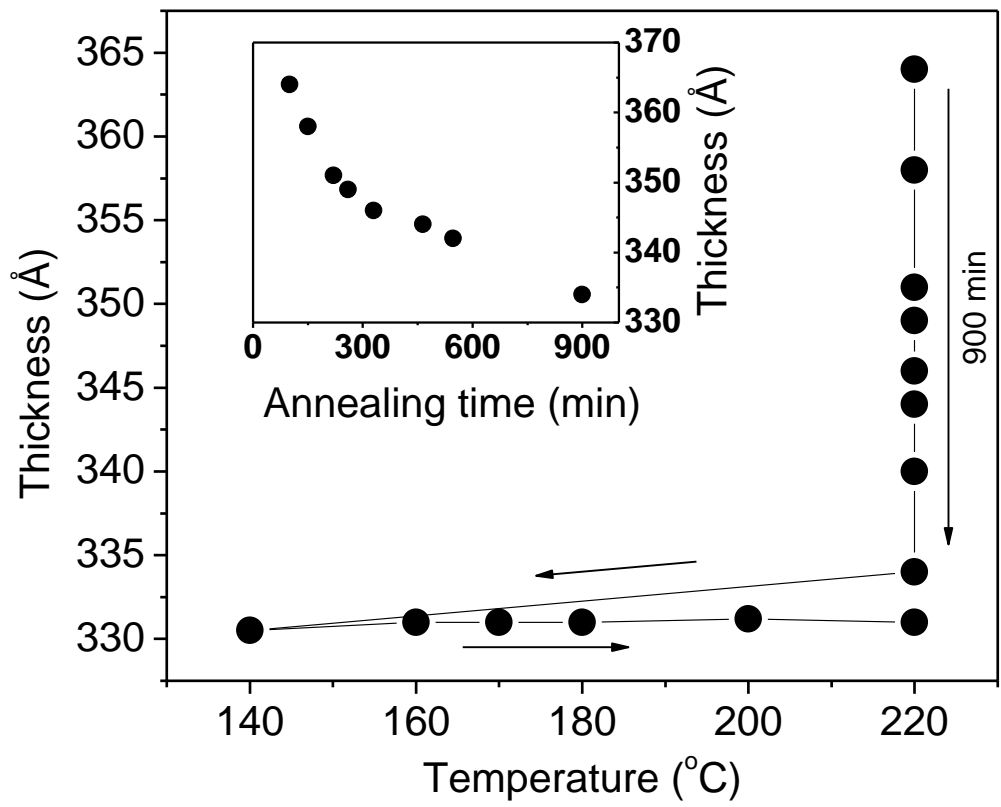

**Figure 3.**

Formation of monocyclic imide from polyacrylamide:

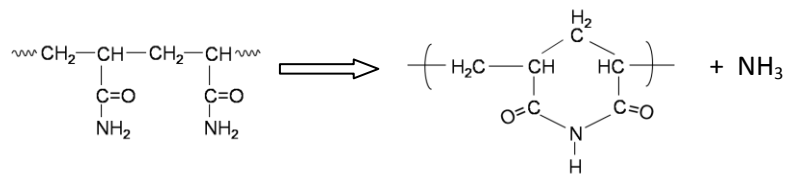

Formation of bicyclic imide from polyacrylamide:

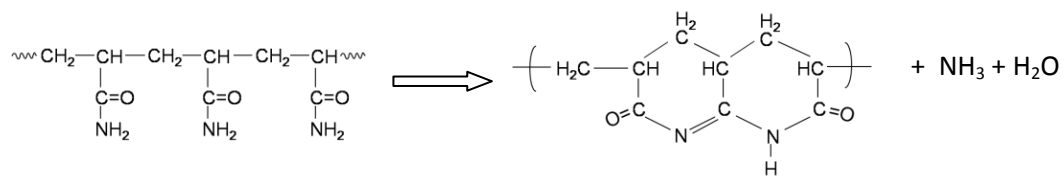

**Figure 4.**

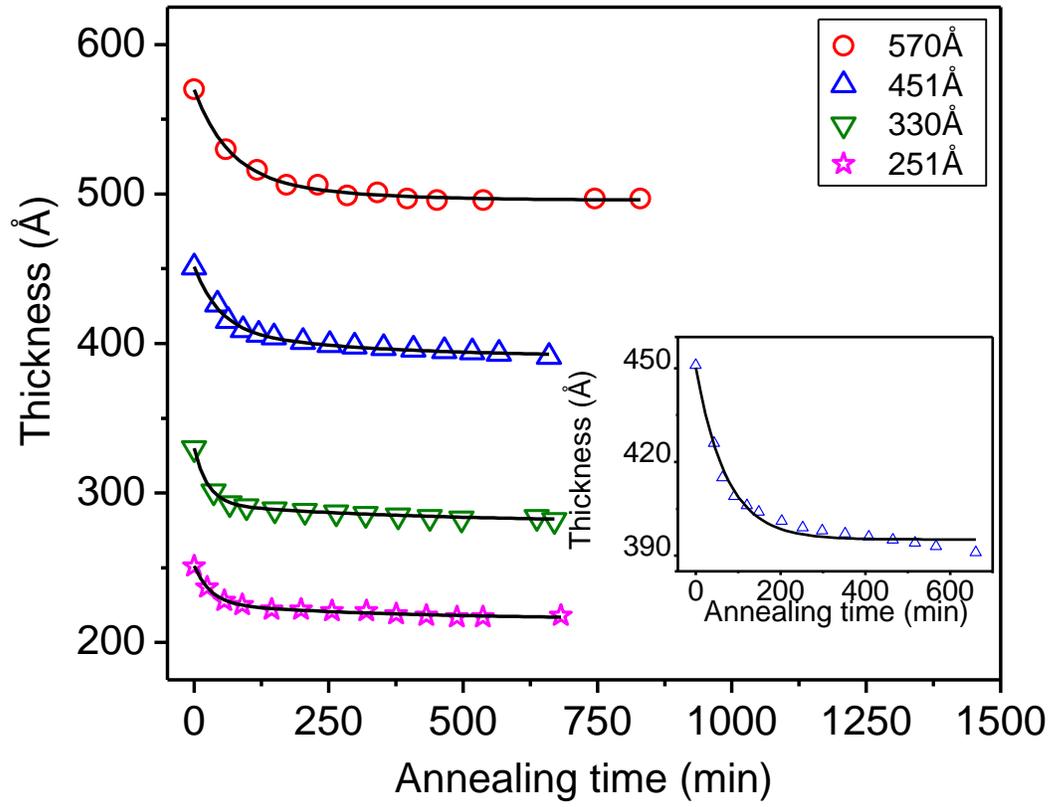

**Figure 5.**

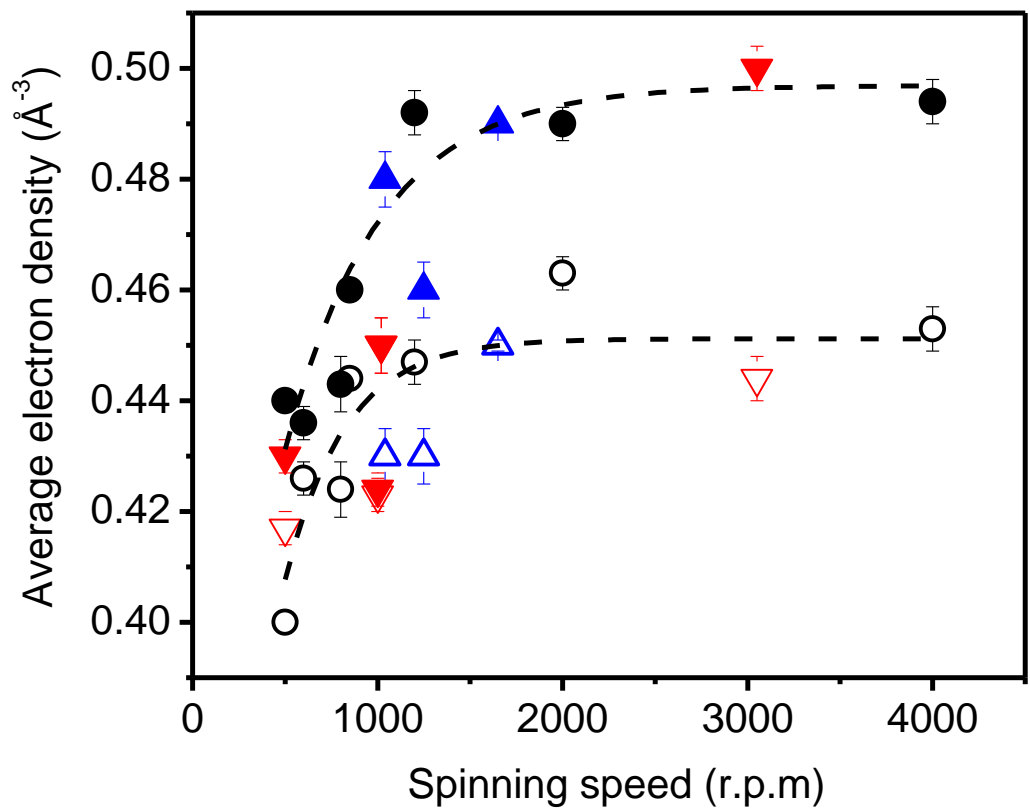

**Figure 6.**

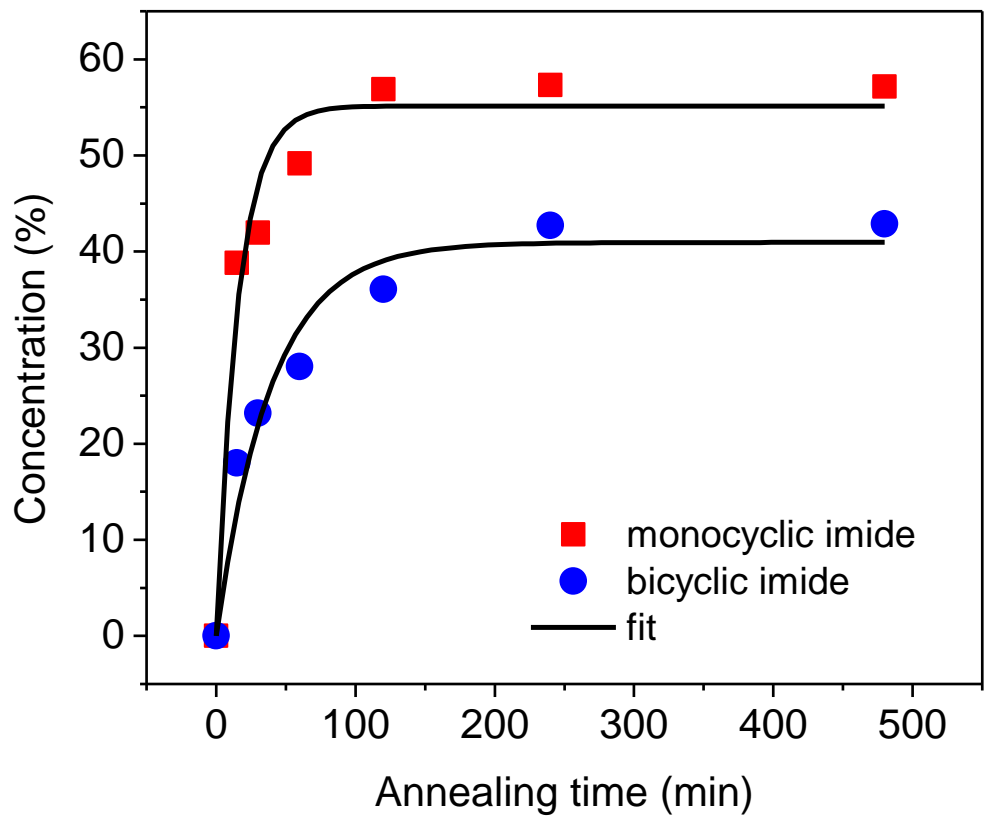

**Figure 7.**